\def\cH{{\cal H}}
\def\cM{{\cal M}}
\def\ket#1{\left|\, #1\right\rangle}
\def\ttm#1#2#3#4{\left(\begin{array}{cc}#1&#2\\#3&#4\end{array}\right)}

\documentclass{article}
\usepackage[hypertex]{hyperref}

\title{Quantum superdense coding {\em \`a la\/} Everett \\ Comparison with teleportation}
\author{George Svetlichny\footnote{Departamento de Matem\'atica, Pontif\'{\i}cia Universidade Cat\'olica, Rio de Janeiro, Brazil \newline
svetlich@mat.puc-rio.br \hfill \url{http://www.mat.puc-rio.br/\~svetlich}}}
\begin{document}
\maketitle

\begin{abstract}We analyze quantum superdense coding as would be seen in Everett's many worlds interpretation of measurement and compare it to Everettian teleportation.
\end{abstract}

\section{Introduction}
In a previous paper \cite{svet:arXiv:quant-ph/0610081} we discussed the quantum teleportation of a qubit in terms of Everett's multiple world interpretation. Here we address the question of quantum superdense coding within the same interpretation. Quantum superdense coding is often discussed jointly with teleportation as there is some superficial similarity between the two processes. By expressing both processes in a unified context, in which all actions  are expressed by unitaries, one can clarify the relation between the two processes. We find that both use control-unitary gates for all information transfers pertaining to one agent, and physical transfer from one of the agents to the other.

As in the previous paper we introduce two agents, Alice and Bob, and assume all the relevant actions take place within an ``Everett bubble'' in which the multiple world ontology holds, but allow that prior and subsequent to the process being analyzed, one can argue within a conventional Copenhagen interpretation for state-preparations or result analyses.

Alice and Bob share an entangled pair of qubits, each one posessing one of the partners. Specifically,
\[\Lambda=\ket0_a\ket0_b+\ket1_a\ket1_b.\]
Here the indices \(a,\,b\) indicate the partners in possession of Alice and Bob respectively.
We use subscripts on kets to also label the Hilbert spaces to which they belong, thus \(\Lambda \in \cH_a\otimes\cH_b\). Our vectors are not necessarily normalized and the identifying subscripts will be omitted when no ambiguity can arise. At times we also omit the tensor product sign \(\otimes\) between ket tensor factors when this helps clarity of expression.

 Alice also has two c-bits' worth of information. Since in an Everett world there is no distinction between classical and quantum, nor a clear notion of what ``knowing some information'' is,  we postulate that prior to the ``Everett bubble'' there is a prepared separated two qubit state in the computation basis \({\ket p}_c{\ket q}_d\in \cH_c\otimes \cH_d\), with \(p,\,q\in \{0,1\}\). This is Alice's two c-bits of ``knowledge'' which once in the Everett bubble she can only manipulate by  applying unitaries to them, or to them tensored with some ancilla.
 Alice can use quantum operations on her ``knowledge" which may seem strange but not illegitimate.

 So the initial state within the bubble is
\begin{equation}\label{bubinit}
\ket p\ket q\otimes \Lambda =\ket p\ket q\ket0_a\ket0_b+\ket p\ket q\ket1_a\ket1_b,
\end{equation}
tensored with any ancilla needed by Alice and Bob within the bubble (only Bob will need them). From now on the Everett ontology holds until Alice and Bob complete their actions.

Denote by \(\sigma_{pq}\), \(p,q\in \{0,1\}\) the unitary matrices:
\[\sigma_{00}=\ttm1001,\quad \sigma_{01}=\ttm0110,\quad\sigma_{10}=\ttm0{-1}{1}0,\quad\sigma_{11}=\ttm100{-1}.\]
When the \(\sigma \) operators act on qubits we represent these  by:
\[\ket 0=\left(\begin{array}{c}0 \\ 1\end{array}\right),\quad\ket 1=\left(\begin{array}{c}1 \\ 0\end{array}\right).\]

Alice acts on Hilbert space \(\cH_c\otimes\cH_d\otimes\cH_a\) by the unitary \(U_a\) defined by
\begin{equation}\label{ua}
\ket p\ket q\otimes{\ket x}_a\mapsto \ket p\ket q\otimes \sigma _{pq}{\ket x}_a
\end{equation}

Note that \(U_a\) is a control-unitary operation where the qbits \(\ket p\ket q\) control the unitary operation applied to qubit \(\ket x\). We designate \(U_a\) therefor by c-\(U_\sigma\) by which it will be thus identified in the appropriate circuit diagram below.

One easily verifies:
\[\sigma _{pq}\ket 0=(-1)^p\ket{p+q}, \quad\sigma _{pq}\ket 1=\ket{p+q+1}\]
where addition within qubit labels is modulo \(2\).

The effect acting by \(U_a\) on (\ref{bubinit}) is to produce the state
\begin{equation}\label{uabubinit}
\ket p\ket q\otimes\{(-1)^p{\ket{p+q}}_a{\ket 0}_b+{\ket{p+q+1}}_a{\ket1}_b\}.
\end{equation}

Alice now physically sends the qubit in \(\cH_a\) to Bob who now performs  a measurement in the Bell basis in \(\cH_a\otimes\cH_b\). In the Everett picture we describe the measruement as a unitary acting on the Hilbert space \(\cH_a\otimes\cH_b \otimes \cH_E\) where the third factor is the Hilbert space of ``pointer positions'' with four basis elements \(\ket{xy}_E\) where \(x,\,y\in\{0,1\}\). We consider \(\ket{00}_E \) as the ``initial position'' of the measuring instrument and consider that this state was present at the beginning of the Everett bubble. The Bell basis is given by
\begin{equation}\label{bellbasis}
\psi _{xy}=\ket x\ket y+(-1)^y\ket{x+1}\ket{y+1}.
\end{equation}

According to the von Neuman paradigm of the measurement process, the unitary \(U_b\) representing the measurement satisfies
\begin{equation}\label{ub}
U_b:\ket{00}_E\otimes\psi _{xy}\mapsto \ket{xy}_E\otimes \psi _{xy}.
\end{equation}

The unitary \(U_b\) is also of the control-unitary type. The Bell basis states \(\psi_{xy}\) control the ``pointer positions" of the ``measuring apparatus" in  \(\cH_E\). We designate this unitary by c-\(U_\cM\) by which it will be thus identified in the appropriate circuit diagram below.

It is easy to verify that state  (\ref{uabubinit}) 1s:

\[\ket{p}_c\ket{q}_d\otimes\psi _{pq}.\]

Tensoring this state with Bob's ancilla (measuring instrument) \(\ket{00}_E\) and performing the ``measurement'' (applying \(U_b\)) one finally has the state:

\begin{equation}\label{bobknows}
\ket{p}_c\ket{q}_d\otimes\{\psi _{pq}\otimes\ket{pq}_E\}.
\end{equation}

Bob's ``instrument{}'' thus points to the label \(pq\) and so he ``knows{}'' now what Alice ``knew{}'' initially. This transfer if {\em two\/} c-bit's worth of knowledge by a transfer of {\em one\/} qubit is what is known as superdense coding.

\section{Teleportation}

We shall want to compare the above procedure to that of teleportation of one qubit in the Everett interpretation. Just as before Alice and Bob share the same entangled two-qubit state \(\Lambda \). Besides this, Alice hods an unknown qubit \(\phi _u=\alpha \ket0_u+\beta \ket1_u\) and a two-qubit ancilla (``measuring instrument'') \(\ket{00}_E\).
The initial state in the Everett bubble is thus:
\begin{equation}\label{telinitbub}
\ket{00}_E\otimes\psi _u\otimes\Lambda .
\end{equation}
She now performs a Bell basis measurement in \(\cH_u\otimes\cH_a\), implemented by a unitary \(U_A\) whose action is identical to that of (\ref{ub}). This results in the state:

\begin{eqnarray}\nonumber
&\ket{00}_E\otimes\psi_{00}\otimes\{\alpha\ket0_b+\beta\ket1_b\} + \ket{01}_E\otimes\psi_{01}\otimes\{\alpha\ket1_b-\beta\ket0_b\}\,+&\\ \label{uainittel} &\ket{10}_E\otimes\psi_{10}\otimes\{\alpha\ket1_b+\beta\ket0_b\} + \ket{11}_E\otimes\psi_{11}\otimes\{-\alpha\ket0_b+\beta\ket1_b\}.&
\end{eqnarray}
We now assume Bob eventually has access to the states in \(\cH_E\) by these being physically transferred to him from Alice  This is the counterpart of the classical signal from Alice to Bob in the usual teleportation protocol. Bob now applies a tripartite unitary operation \(U_B\) acting on \(\cH_E\otimes\cH_b\)
and defined by
\[U_B:\ket{xy}_E\ket z_b\mapsto(-1)^{y(z+1)}\ket{xy}_E\ket{z+x+y}_b.\]
This results in the final state
\[\left(\sum_{xy}\ket{xy}_E\otimes\psi_{xy}\right)\otimes\phi_b\]
where \(\phi_b=\alpha\ket0_b+\beta\ket1_b\). Thus an exact copy of Alice's unknown state \(\phi_u\) is now in Bob's possession as one of the factors of a separated pure state.

\section{Circuits}

To fully appreciate the relationship between the two processes we give below the quantum circuits as they function in the Everett bubble. These are given below, superdense coding on top, teleportation bottom:

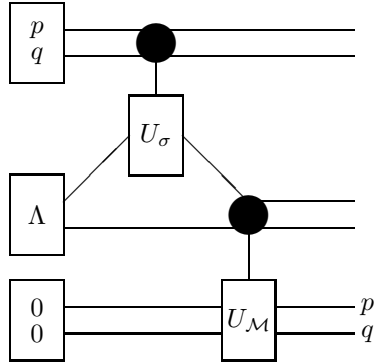
\begin{center}
\begin{picture}(150,140)
\put(0,105){\framebox(20,30){}}
\put(10,125){\makebox(0,0){\(p\)}}
\put(10,115){\makebox(0,0){\(q\)}}
\put(0,40){\framebox(20,30){\(\Lambda\)}}
\put(45,70){\framebox(20,30){\(U_\sigma\)}}
\put(55,120){\circle*{20}}
\put(55,100){\line(0,1){20}}
\put(20,125){\line(1,0){110}}
\put(20,115){\line(1,0){110}}
\put(20,60){\line(1,1){25}}
\put(20,50){\line(1,0){110}}
\put(65,85){\line(1,-1){25}}
\put(90,55){\circle*{20}}
\put(90,60){\line(1,0){40}}
\put(90,50){\line(0,-1){20}}
\put(80,0){\framebox(20,30){\(U_\cM\)}}
\put(0,0){\framebox(20,30){}}
\put(10,10){\makebox(0,0){\(0\)}}
\put(10,20){\makebox(0,0){\(0\)}}
\put(20,20){\line(1,0){60}}
\put(20,10){\line(1,0){60}}
\put(100,20){\line(1,0){30}}
\put(100,10){\line(1,0){30}}
\put(135,20){\makebox(0,0){\(p\)}}
\put(135,10){\makebox(0,0){\(q\)}}
\end{picture}
\end{center}

\vskip 32pt

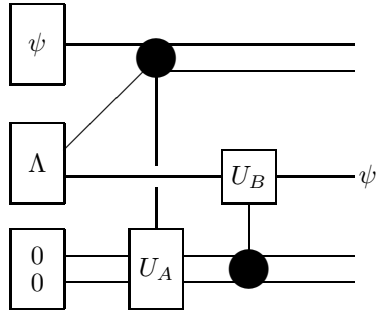
\begin{center}
\begin{picture}(140,115)
\put(0,85){\framebox(20,30){$\psi$}}
\put(0,40){\framebox(20,30){\(\Lambda\)}}
\put(45,0){\framebox(20,30){\(U_A\)}}
\put(55,95){\circle*{20}}
\put(55,90){\line(0,-1){36}}
\put(55,46){\line(0,-1){16}}
\put(55,90){\line(1,0){75}}
\put(20,100){\line(1,0){110}}
\put(20,60){\line(1,1){30}}
\put(20,50){\line(1,0){60}}
\put(100,50){\line(1,0){30}}
\put(135,50){\makebox(0,0){\(\psi\)}}
\put(90,15){\circle*{20}}
\put(90,40){\line(0,-1){20}}
\put(80,40){\framebox(20,20){\(U_B\)}}
\put(0,0){\framebox(20,30){}}
\put(10,10){\makebox(0,0){\(0\)}}
\put(10,20){\makebox(0,0){\(0\)}}
\put(20,20){\line(1,0){25}}
\put(65,20){\line(1,0){65}}
\put(20,10){\line(1,0){25}}
\put(65,10){\line(1,0){65}}
\end{picture}
\end{center}

\begin{itemize}
  \item In both situation there are two control-\(U\) gates, one with two qubits controlling one and one with two qubits controlling two.
  \item In both situations there is a physical transfer of quantum states from Alice to Bob. In teleportation this is a pair of separated qubits encoding two c-bit. In superdense coding it is one quibit, Alice's part of the entangled state.
  \item In both ``information" is passed through a control-\(U\) gate from states held by Alice to the state which is physically transported to to Bob.
\end{itemize}

\section{Conclusions}
Placing all classical dynamics, especially communication, as unitary transformation in the Everett picture brings out certain universal features of quantum processing. All local transfer of information can be done by appropriate control-unitary gates. This is in keeping with results proved in \cite{gott-chua:Nature402.390,svet:arXiv:quant-ph/0601093} concerning the universality of a teleportation-type protocol. Physical transport of states is also necessary here, as is probably the case in any transfer of quantum information between distant agents.\footnote{Contrafactual quantum communications \cite{noh:ptl103.230501} may seem to belie this, but Alice and Bob still need to communicate classically, which in the Evertett picture corresponds to physical transfer of states.}

\section{Acknowledgements}
This research was partially supported by the Conselho Nacional de Desenvolvimento Cient\'{\i}fico e Tecnol\'ogico (CNPq), and the Funda\c{c}\~ao de Amparo \`a Pesquisa do Estado do Rio de Janeiro  (FAPERJ).

\end{document}